# A Systematic Review of the Digital Interventions for Fighting COVID-19: The Bangladesh Perspective

MUHAMMAD NAZRUL ISLAM[1] AND A.K.M. NAJMUL ISLAM[2, 3]
[1] Department of Computer Science and Engineering, Military Institute of Science and Technology, Mirpur Cantonment, Dhaka-1216, Bangladesh
[2] Department of Future Technologies, University of Turku, 20500 Turku, Finland
[3] LUT School of Engineering Science, LUT University, 53850 Lappenranta, Finland

Corresponding author: Muhammad Nazrul Islam (e-mail: nazrul@cse.mist.ac.bd)

**ABSTRACT** The objective of this paper is to synthesize the digital interventions initiatives to fight against COVID-19 in Bangladesh and compare with other countries. In order to obtain our research objective, we conducted a systematic review of the online content. We first reviewed the digital interventions that have been used to fight against COVID-19 across the globe. We then reviewed the initiatives that have been taken place in Bangladesh. Thereafter, we present a comparative analysis between the initiatives taken in Bangladesh and the other countries. Our findings show that while Bangladesh is capable to take benefits of the digital intervention approaches, tighter cooperation between government and private organizations as well as universities would be needed to get the most benefits. Furthermore, the government needs to make sure that the privacy of its citizens are protected.

## I. INTRODUCTION

The '2019 novel coronavirus (2019-nCoV)' or 'COVID-19' is a new strain of coronavirus that is linked to the same family of viruses such as Severe Acute Respiratory Syndrome (SARS). The outbreak of COVID-19 was first identified in Wuhan, China in December 2019, which then spread to the rest of the world. As of today (06 April 2020), coronavirus has spread in 208 countries and territories affecting over a million people, and 69,501 deaths [1]. Figure 1 shows the total infected cases of COVID-19 around the world [1] and it has been rapidly rising. Mathematical modeling suggests that one in three people given hospital treatment for coronavirus will need intensive care [2]. Therefore, it is expected to overwhelm the heath-care services of many countries.

Bangladesh is a densely populated country with around 170 million people, and a population density of 1265 persons per square kilometer [3]. The first three cases of corona affected patients in Bangladesh were reported by the country's Institute of Epidemiology, Disease Control and Research (IEDCR) on 8th March 2020 [4]. As of 02 April 2020, according to the IEDCR report, a total of 123 persons are affected, 13 died and 33 have recovered; 66,071 persons are in home-quarantine and 420 are in isolation [5]. Figure 2 shows the total infected cases of COVID-19 in Bangladesh [5].

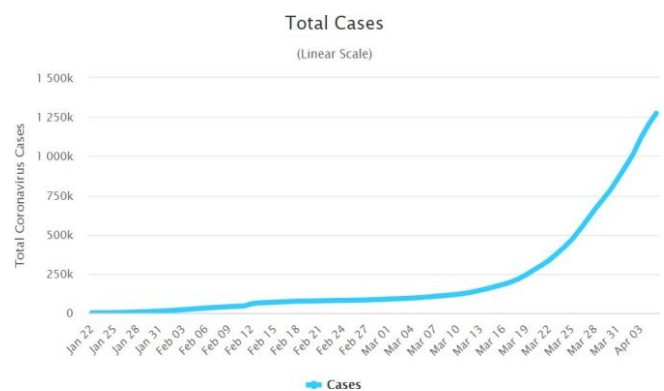

**FIGURE 1.** Infected cases of COVID-19 around the world [1]

The pandemic spread of COVID-19 has brought many countries to a halt. Academic institutions and businesses were closed down; houses, villages, even the entire cities were quarantined to reduce the spread of COVID-19. People are panicked due to the mass information (and misinformation) on COVID-19 on social media and other online sources [6]–[8]; and the healthcare systems of different countries are crumbling under the encumbrance. The government of Bangladesh and private organizations have taken a number of initiatives to fight against the pandemic spread of COVID-19 [9]. WHO declared declared COVID-19 a global pandemic on 11th March, 2020 [10]. Consequently, Bangladesh government imposed total lockdown and deployed armed forces to assist civil administration from 26th March, 2020. The educational institutions were closed down from 18th March, 2020 [11]. However, considering the fast-spread of the novel corona virus and its divesting examples throughout the whole world as shown in Figure 2; and considering the population density, health illiteracy, socio-economic situation and the health services of Bangladesh, it would be tough for Bangladesh to fight against the pandemic spread of novel corona virus. Therefore, digital solutions to provide health care services were deemed important. Digital interventions or the digital health interventions can play a vital role to provide health care services for the containment of the pandemic spread of COVID-19 as observed in many developed countries [12] [13] [14]. The term digital health intervention is defined by the WHO [15] as "a discrete functionality of digital technology that is applied to achieve health objectives and is implemented within digital health applications and ICT systems, including communication channels such as text messages".

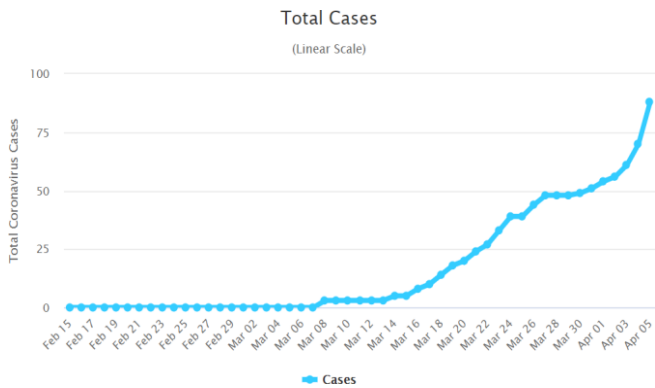

**FIGURE 2.** Infected cases of COVID-19 in Bangladesh [5]

Although several epidemic and pandemic diseases occurred in the recent past like Ebola Virus Disease (EVD) and H1N1 flu (swine flu) [16], COVID–19 pandemic can be differentiated based on the number of infected countries, cases and applied digital interventions all over the world. Therefore, there is limited knowledge on how digital interventions can help in the containment of a global pandemic. Recent reports suggest that different kinds of digital solutions are being used by the different countries in order to combat the pandemic spread of COVID-19. These technological interventions are published in various websites and newspapers in a very scattered way [13] [17] [18] [19]. Therefore, a research is needed to systematically gather these initiatives and summarize them for communicating it to the academic society as well as other relevant stakeholders such as governments, non-governmental agencies, ICT companies, among others. Developing countries like Bangladesh have also taken a number of ICT-based initiatives to fight against the pandemic spread of the novel corona virus. However, what kinds of digital initiatives have been taken and how do these initiatives differ from the developed world are required to be studied to plan for future digital intervention strategies. Thus, it is utmost important to understand what kind of digital solutions are used to combat the pandemic spread of cornonavirus; and then highlight the possible opportunities and the challenges to adopt such digital interventions in the context of a developing nation like Bangladesh.

Therefore, the objective of this study is to explore the digital intervention for the containment of pandemic spread of COVID-19 across the globe as well as in the context of Bangladesh. The outcome of this research will greatly contribute to understand what kind of technology could be used and for what purposes. The findings may help the developing nations like Bangladesh to take further necessary initiatives related to digital interventions to measure, prevent and control the pandemic spread of COVID-19.

## II. RESEARCH METHOD

As the topic is still emerging and there are no academic literature available in this area, we followed an online content review approach to attain our research objectives. We note that online content review approach is widely accepted in research, and popularly known as digital ethnography, online ethnography and netnography [20], [21]. Google search engine was used to find the related online content. The online content were analysed through systematic coding and interpretation.

The search strings used to find the available online content from a global perspective were "digital intervention and corona virus", "digital intervention and COVID-19", "information technology and corona virus", "information technology and COVID-19", "digital services and corona virus", "digital services and COVID-19", "artificial intelligence and corona virus", "artificial intelligence and COVID-19", "mobile application and data science and robot and corona virus", "mobile application and data science and robot and COVID-19", "electronic health services and corona virus", and "electronic health services and COVID-19". The similar set of keywords appended with 'and Bangladesh' were also used to find the online content explicitly focusing to the context of Bangladesh, for example, "digital intervention and corona virus and Bangladesh", "digital intervention and COVID-19 and Bangladesh", and the likes.

The search results produced more than 1000 sites of contents. We then removed the online content that discussed the same topic, especially the news articles that focus on the same digital/ICT interventions but published in multiple newspapers; online contents that are not focusing to our research objectives; and the contents written other than English and Bengali. After applying this inclusion-exclusion process, we finally selected 57 online resources that include electronic news article, article written as a blog, online press releases, and the websites/web contents of different organizations. The flowchart of inclusion and exclusion process to select the online resources or electronic resources for content analysis is shown in Figure 3.

The selected article were reviewed systematically for coding (after extracting related data) and interpretation. This review study primarily extracted two types of data: the technology or digital solution used in pandemic spread of COVID-19; and the purpose of using such digital technologies for the containment of pandemic spread.

Then the data were synthesized and analyzed to explore the digital interventions from a global perspective and in context of Bangladesh. Finally, we conducted a comparative analysis between these two sets of findings to understand the underlying facts of digital interventions in context of Bangladesh and future perspectives of adopting any new digital interventions in Bangladesh to combat the pandemic spread of COVID-19.

## III. DIGITAL INTERVENTION AND COVID-19 PANDEMIC

This section summarizes digital interventions during the pandemic spread of COVID-19. Our review found many major ICT interventions around the globe [12]–[14], [22], [23]. We found automatic vehicles have been used to reduce workload of medical staffs and avoid cross-infection in China. For example, two White Rhino unmanned vehicles were used in Wuhan to supply medical staffs, medicines and foods [24]. The use of drones for different purposes was observed in several countries including China, Italy, Singapore, South Korea, France, and USA [25], [26]. Drones are used to deliver medical supplies (e.g., thermal imaging camera) and patient samples to improve virus detection. Agricultural drones are used to spray disinfectants in the countryside. Some countries used drones to ensure the lockdown. Drones that are equipped with the face recognition capability are used to broadcast warnings for staying at home and wearing protective facemasks.

The use of robots is also noteworthy to mention during the pandemic spread of COVID-19 [27]–[30]. Robots are used in different capacity in different countries such as USA, China, Tunisia, Greece, Spain, and Italy for measuring body temperature, recording other relevant data, spraying disinfectants and cleaning, assisting patient diagnosis, monitoring the quarantine patients, preparing meals at hospitals, serving foods in restaurants, vending foods, and dispensing hand sanitizers, among others [17], [31]. Virtual humans are used to support nurses and doctors to screen sick patients, monitor patient recovery or help hospital staffs to answer calls or patient questions [17].

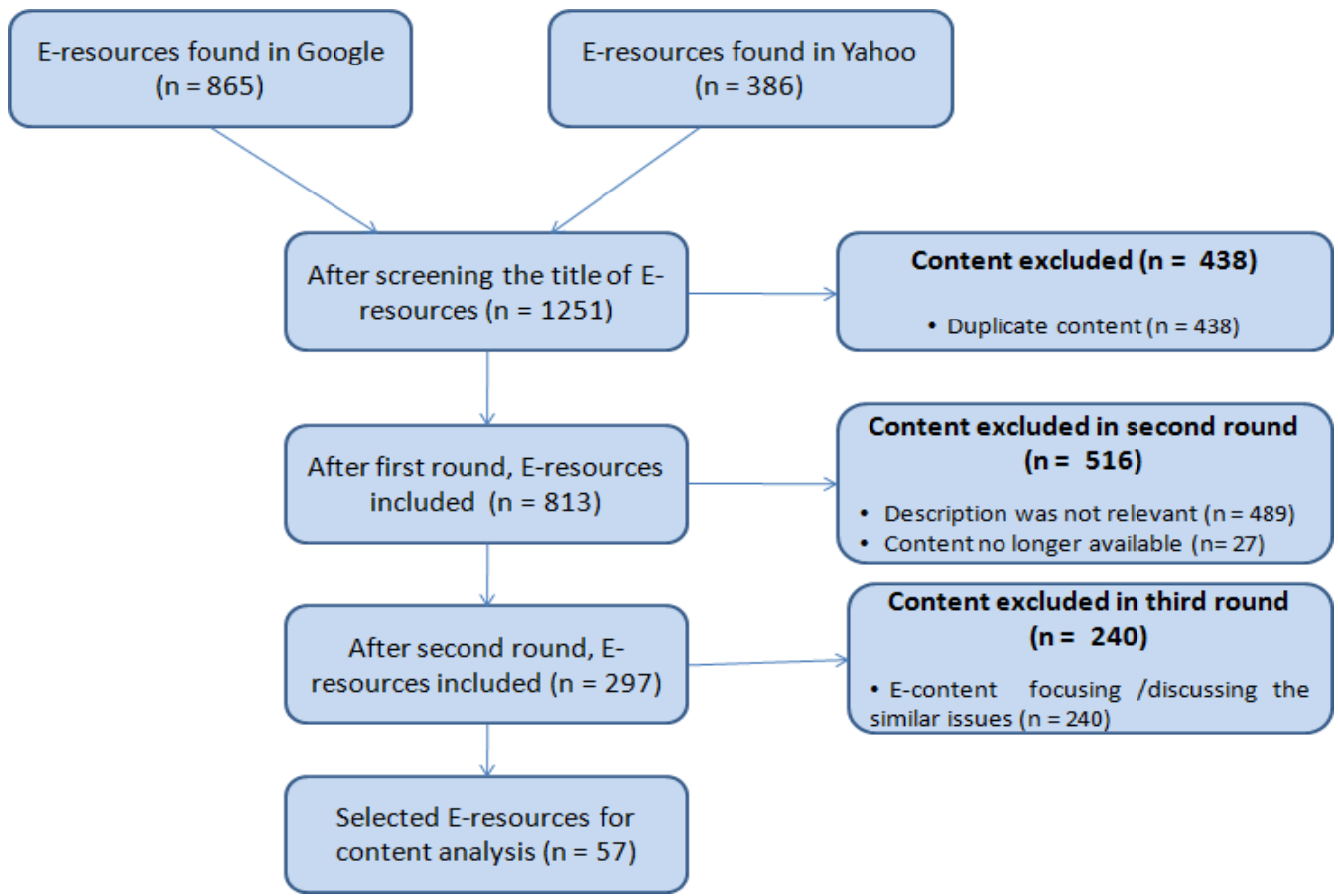

**FIGURE 3.** Online resources inclusion and exclusion process flowchart

Artificial intelligence (AI) is being used to help in the crises of COVID-19 pandemic in a number of ways like predicting affected patients, maintaining social distance, and screening the patient risk in Singapore, South Korea, China, Taiwan, USA, UK, and the EU countries [32]–[35]. AI is used to detect people who are not using the protective mask. It is also used to diagnose or predict coronavirus affected persons through facial recognition technology, interpret radiological image, provide intervention recommendations, and monitor temperature, among others [19]. AI is also heavily used in online screening and consultation platforms, for example, an AI based coronavirus screening system is developed that can start automatic video conversation if a particular person is detected as a suspected patient. Another AI based system first discusses with the patients through text messages before transferring them to the doctors.

Big data and data analytics were used for collecting, sorting and analysing the huge masses of data produced on the Internet and other sources in Malaysia, USA, Italy, China, Singapore, and UK, among many other countries [36]–[38]. These techniques are used to build predictive models and turning point projection to prevent and control the pandemic. Data science and AI together are used to identify, track and forecast outbreaks, diagnose the virus, process healthcare claims, develop drugs, and develop advanced fabrics for protection [19]. For example, a big data based health monitoring system was developed in China that uses data related to the individuals travel history, time spent in virus hotspots, and close contact with any affected people to identify and assess the risk of each citizen [19]. Furthermore, supercomputing is used for vaccine and drug development [39].

Internet of Things (IoT) is used to track clinical specimens. For example, in Italy, an IoT based dash clicker ('Nurse Button') was used to connect patients and doctors in a field hospital with 5,500 hospital beds and 500 intensive care beds [40]. Wearable technologies are used to provide telemedicine services, crowd activity monitoring, and care in quarantine area [19], [41]. For example, a wearable device that can be placed on throat to monitor temperature, cough and respiratory activity, has been developed in the US. It is currently used in a limited scale to monitor the COVID-19 patients from hospital. In China, sensor technology is used for remote health monitoring [41], [42].

Tele and video consultation services have also been used during the pandemic in all the infected countries. For example, national health service of England strongly recommended to use video-based consultation to help and reduce the risk of spreading the coronavirus [43]. Chatbots are used to share information, reassure people, help to assess COVID-19 risk, help patients to get treatment and keep them away from emergency care centers [44].

Online assessment/screening systems are used to allow individuals to assess the possibility of being a carrier of COVID-19 in all the infected countries including China, Singapore, USA, and EU countries [45], [46]. Furthermore, internet, websites, mobile applications and social network sites were used to monitor the disease outbreak and information dissemination [17]. For example, an online resource hub that includes blog posts, webinars, and other useful links for helping users to manage stress, staying active, eating healthy diets and sleeping well for dealing with COVID-19 pandemic [44]. China used a smartphone app with color codes (green, yellow or red) to find out people who should be allowed in public spaces or should be quarantined [18]. Similarly, South Korea developed the *Corona 100m app* that alerts users when they come within 100 metres of a COVID-19 patient [26]. Interactive maps are used to provide the information related to the surveillance of the epidemic. Virtual Reality (VR) is used to teach kids about coronavirus safety [17].

Figure 4 summarizes the digital interventions to fight with the pandemic spread of COVID-19.

## IV. DIGITAL INTERVENTION TO COMBAT COVID-19 IN BANGLADESH

As of Feb 2020, according to the Bangladesh Telecommunication Regulatory Commission (BTRC) there are approximately 100 millions Internet users, 166 millions mobile phone subscribers [47] [48]. Due to the wider adoption of mobile phones and Internet, it deemed possible to use digital interventions during the COVID-19 pandemic. In response to pandemic spread of COVID-19, several initiatives related to the digital technologies are taking place in Bangladesh. The existing initiatives include the following.

i National Corona Portal-Corona Info: This is the national web portal in Bangladesh for providing COVID-19 related information and resources [5]. The portal provides current status of the total number of confirmed cases, sample being tested, total recovery, and total death due to COVID-19, among other information. It also provides information related to the total number of quarantined people, people at isolation and the number of people being released from quarantined or isolation. The corona symptoms, WHO guidelines, list of emergency numbers, safety and prevention tips, video messages from key personnel about public awareness, recent news and digital press releases related to corona, are updated here. Apart from these, doctors and interested persons may apply to contribute as health workers and to participate training on COVID-19.

ii Information chat bots: The 'COVID-19 Response Bangladesh', 'Mita Chatbot' and the 'Bloodman' chatbots are embedded in Facebook Messenger, and a few other chatbots like the 'eGeneration Corona Bot' and the 'Corona INFO BOT' are hosted at national Corona portal [5], [49]. The chatbots can be used both in Bengali and English. The chatbots source information from IEDCR,

WHO, and CDC. The chatbots offer COVID-19 related information, advice on mental health, doctor consultation, and access to emergency, among others.

iii Electronic and mobile awareness initiatives: All major mobile phone operators started a number of awareness and preventive initiatives among their customers [50]. For example, as part of an awareness drive, Grameenphone sent awareness and preventive SMSs on coronavirus pandemic to its subscribers, provided cautionary messages such as "Stay Home" next to the signal bar on the phones. It also started a call service (free of charge) to *3332 to discuss and get information on COVID-19 [51]. Ministry of Health and Family welfare of Bangladesh government also started a national COVID-19 surveillance system to assist the people of Bangladesh in the pandemic spread of novel corona virus [52].

iv Interactive Voice Response (IVR) services: The BTRC directed all mobile operators in Bangladesh to start a 90-second call under interactive voice response service [53]. During each IVR session users will be asked several questions including their age, whether they have fever or cough, breathing problems, if they came in contact with any corona virus infected person or someone who recently returned from abroad.

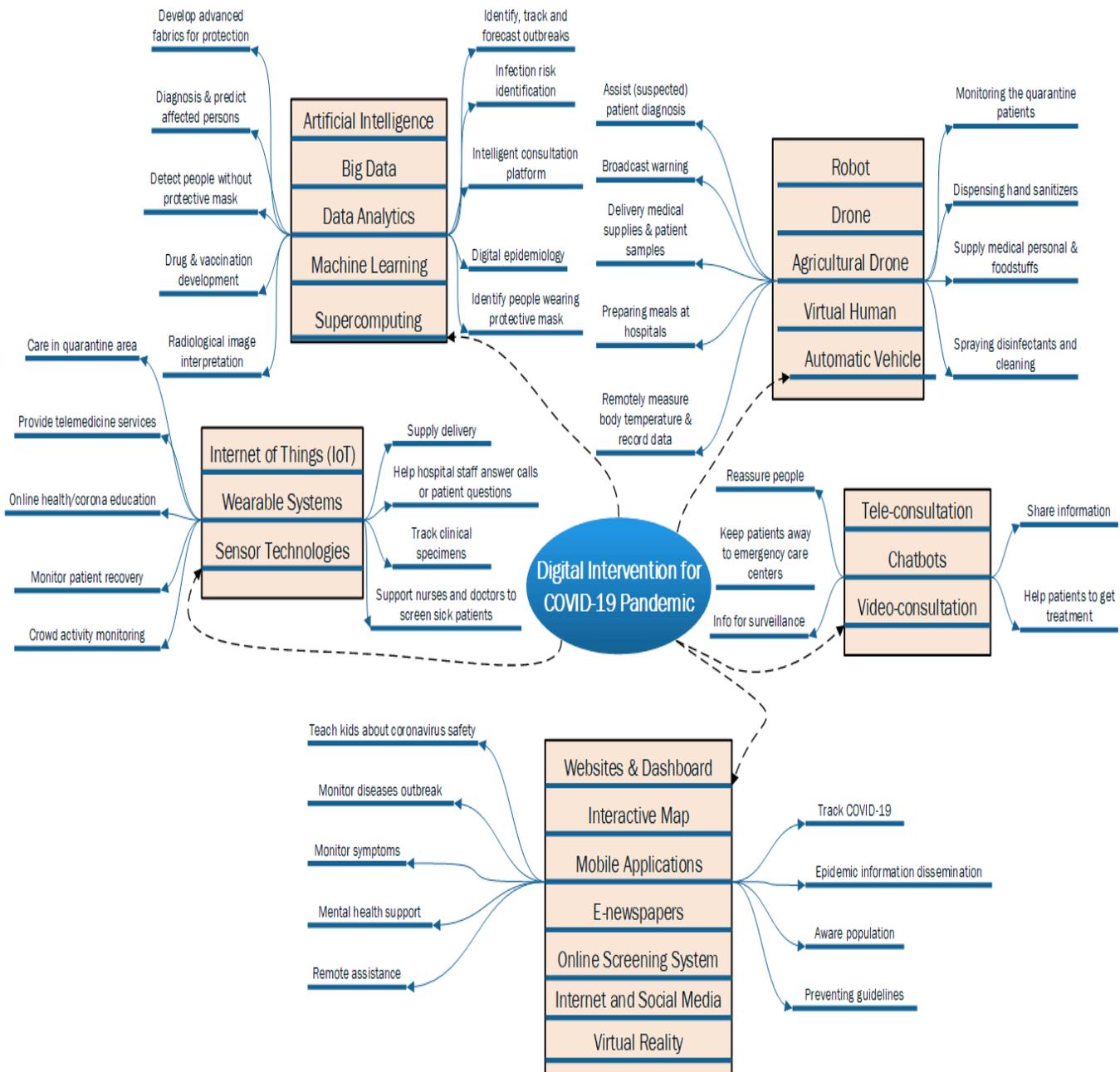

**FIGURE 4.** Digital intervention to fight with the pandemic spread of COVID-19

v  COVID-19 self-assessment: A web service available at [54] is released to check the corona symptoms in Bengali. The system takes all necessary information about health condition and any close contact with corona affected people, and person recently return from abroad for assessment. Few other corona testing services like 'Corona survey', 'COVID-19 Checker', 'COVID-19 Self Test', and 'Test Corona' are developed by different non-government organizations, and are hosted at the national Corona portal [5].

vi  Mobile Applications: At present, only a few number of mobile applications are developed in Bangladesh [5]. The apps are mainly developed to provide preventive guidelines, symptoms of corona virus, awareness information, information for emergency care, self-screening to check if anyone is infected or have the symptoms of corona infection, track the quarantine people, provide the current status of infected, recovery and death, and the likes. The Table 1 provides an overview of the existing apps and their purpose of development.

vii  Social media/Facebook page: As Facebook and YouTube are popular in Bangladesh, many Facebook pages are created to provide the awareness information and up-to-date corona status to the citizens. Several video documentaries have been prepared in Bengali and released through YouTube and Facebook to make it accessible to mass population for making people aware and prevent corona virus. Corona related press conferences and other awareness programs are organized thorough Facebook live in a regular manner.

viii  Preventive, awareness and controlling initiatives: A number of initiatives to educate and aware people about corona virus, its prevention and controlling mechanism are taken. Such initiatives include online help desk, dedicated section of a website, COVID-19 dashboard, national guidelines and policies for preventing corona virus, and dedicated window for the e-newspapers, among others by different government and private organizations as well as universities [59]–[61]. For example, SciTech Academy has developed a educational portal on COVID-19 for the Kids (age 5 to 12) to drive children's awareness to deal with the pandemic spread of novel corona virus without being panicked [62].

ix  AI and Big data based predication system and other ICT initiatives: A team from Daffodil International University, Bangladesh developed an AI based corona prediction system that can diagnose COVID-19 from X-ray results. However, the system has not been clinically tested yet to make it usable in the current situation [63], [64]. Another organization (eGeneration) developed a similar system using neural network to analyse the chest X-Ray images to detect COVID-19 and usual pneumonia [65]. Few other institutes are conducting similar kind of research but yet not developed anything for the clinical use. For example, a cost-effective robot is developed by the Sonargaon University, Bangladesh to support the COVID-19 patients and assist doctors for treating corona infected patients [66]. This robot still is in the experiment phase and yet to be deployed in the hospital. Based on the mobile users' self-reported health information (collected through free SMS) and big data analytics, the government is in the process of developing a digital map to track coronavirus cases and find out areas susceptible to contamination [67], [68]. Bangladesh police has used drones in a very limited scale to track and monitor citizens in Chattogram and Satkhira districts to enforce social distance and the lockdowns [69], [70]. Furthermore, the government is sponsoring to arrange a competition to design a cost-effective and efficient ventilator, and other IT-based application that can serve corona infected patients [71].

## V. DISCUSSIONS

In this section, we describe a comparative analysis of the approaches that have been taken by Bangladesh in compare to the rest of the world described earlier.

First, we observe that Bangladesh has been mostly relying on the self-reported data using the previously discussed IVR and self-assessment web portal. This makes sense as testing all people (170 millions) for detecting COVID-19 is practically impossible due to the costs and other practical limitations. In fact, we observe that even developed countries are not able to test all of their citizens. The true infection rate is expected to be much higher than what we see as the confirmed reported number of cases. Therefore, it is understandable why Bangladesh moved to self-reported data approach. However, it has introduced some unique challenges. The self-reported data are subject to biases and consist of noises. Despite the noises, the services are expected to give some benefits such as identifying possible affected areas and quickly allocate resources in the areas or even locking down the entire areas. In general, we observe preliminary initiatives in Bangladesh on using data science and AI approaches. However, in our view it has introduced the challenge of data privacy. This is because there is no comprehensive data privacy protection framework like the GDPR in Europe and the current digital security act of Bangladesh has some loopholes [72], [73]. Most of the digital interventions have been based on the cooperation between public and private sectors. Therefore, citizens' data was shared with the private companies. This leads citizens to worry about their data privacy. However, the government has assured that it will make the data protection act as soon as possible. Many people also believe that if they are identified as corona patients, they will be left out of the society. Therefore, they are reluctant to share information, even they are having corona symptoms.

Second, most of the initiatives have been taken to share information related to COVID-19 in Bangladesh. The information includes statistics on the amount of infected cases, news, and mitigation approaches, among others. Therefore, these initiatives relate to raising awareness among the people about COVID-19. In contrast, countries like China and South

TABLE 1. List of mobile applications developed in Bangladesh for COVID-19 pandemic

| Apps Name & Ref | Platform | Cost | Language | Purpose of the Mobile Application |
| --- | --- | --- | --- | --- |
| Corona Identifier [55] | Android | Free | Bengali | Provide updated corona status, preventive information, emergency information, and notify if any corona affected person comes close (2-3 meter) to the app user. |
| Corona BD App [56] | Android | Free | Bengali | Provide corona status and information related to the prevention, awareness, and emergency contact information |
| Corona Prevention [57] | Android | Free | Bengali | Provides prevention guidelines, symptoms of corona virus, how the virus spreads and tips on emergency care. |
| Quarantine Tracker [57] | Android | Free | Bengali | Assists to track quarantine people using the global positioning system and face detection system. |
| COVID-19 State [58] | Android | Free | Bengali | Provide the updated status mainly about the number of individuals who are infected, recovered, died, under quarantine, and isolation in Bangladesh. |
| Corona Info App [5] | Android | Free | Bengali | This is the mobile version of the national corona portal of Bangladesh, and provides the similar services through mobile platform. |
| Corona Tester App [5] | Android | Free | Bengali | Provide COVID-19 risk assessment service, and information related to COVID-19 prevention, control and treatment. |
| Corona Info - Self Reporting App [5] | Android | Free | Bengali | Individuals may report their health conditions (symptoms related to COVID-19) to get the required treatment services. |

Korea have been using wearable and other related technologies to track people's activities to make sure the infected people are maintaining the quarantined protocol. However, this again introduces the challenge of data privacy that we discussed before.

Third, we note that we did not find much digital intervention approaches regarding psychological well-being across the globe. COVID-19 is not only impacting our physical health but also creating fear among us. Therefore, more interventions would be required to maintain people's psychological well-being during this pandemic.

Fourth, a major challenge from both global and Bangladesh perspectives is to deal with misinformation on online sources [6]. Misinformation creates panic among people and also it affects on people's behavioral change during pandemic. Therefore, we believe this is one potential area where digital intervention may be useful.

Fifth, tele and video consultation services are needed to be deployed in each hospital in Bangladesh as these services were shown effective in this context [74]–[76]. The use of tele-consultation has been proved useful in the global perspective [77] as well. We observed several initiatives by some hospitals and organizations in Bangladesh so far that are providing telemedicine services during the COVID-19 pandemic situation to the citizens of Bangladesh as well as for the Bangladeshi citizens living in abroad [76]. However, in order to increase the coverage area, more hospitals need to start tele-consultation services, and preferably video consultation.

Finally, only a few mobile applications have been developed for coronavirus so far. We note that prior research found that most mobile health applications developed in Bangladesh suffer from usability problems [78]–[82]. Prior works suggest that usability is the key for wider adoption and continued use of services [83], [84]. We did not test the usability of the COVID-19 mobile applications in our research. However, we observed that most of the applications developed for COVID-19 provide similar type of services as well as significant number of citizens are not well aware about these applications. Furthermore, the availability of 4G technology, internet connectivity, and user behaviour towards accessing the mobile internet and mobile applications in Bangladesh clearly depict the opportunity to introduce more useful and usable mobile applications for the containment of the pandemic spread of the coronavirus. Different services such as remote assistance, mental health support, awareness information, education on corona virus, and monitoring quarantine people and communicating with them could be provided through mobile applications. Developing such kind of applications would be a cost-effective, efficient and feasible digital intervention to combat COVID-19 pandemic to a greater extent.

## VI. CONCLUSION

In this paper, we have systematically reviewed the digital intervention initiatives that have been taken across the globe as well as by Bangladesh for the containment of pandemic spread of COVID-19. Based on our comparative analysis, we propose some areas where Bangladesh can focus to reduce the pandemic spread of COVID-19. In summary, we think that some of the digital interventions that have been taken and proved useful in the global perspective may be possible in Bangladesh. In order to achieve this, we think a cooperation between government and private sectors will be needed.

Our study has a number of limitations, but at the same time provides some avenues for future research. First, we used some specific keywords to search for the relevant materials. Therefore, it may be that we missed some important materials that did not emerge from our search queries. Second, it may be that the materials that we have used may have some biases in their reporting. Third, we think timely and up-to-date data are the key to most of the digital interventions that we have identified in this paper. Therefore, future works would be needed to collect accurate data using multiple methods and not just relying on testing facilities. Fourth, while many initiatives have been taken to collect data through digital surveillance, we think that there is a need to develop privacy friendly services. Finally, in this paper we present a comparative analysis between the initiatives taken in Bangladesh and the other countries in the world. Future research can be conducted to present comparisons among other countries.

This article has been accepted for publication in a future issue of this journal, but has not been fully edited. Content may change prior to final publication. Citation information: DOI 10.1109/ACCESS.2020.3002445, IEEE Access

M.N. Islam et al.: A Systematic Review of the Digital Interventions for Fighting COVID-19: The Bangladesh Perspective

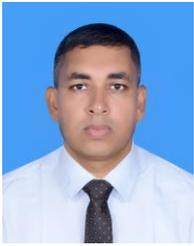

DR. MUHAMMAD NAZRUL ISLAM is an Associate Professor at the department of computer science and engineering at the Military Institute of Science and Technology (MIST), Mirpur Cantonment, Dhaka, Bangladesh. He was awarded a Ph.D. in Information Systems from Åbo Akademi University (Finland) in 2014, a M.Sc. in computer engineering from Politecnico di Milano (Italy) in 2007, and a B.Sc. in computer science and information technology from Islamic University of Technology (Bangladesh) in 2002. Before joining MIST, he was working as a visiting teaching fellow at Uppsala University, Sweden and as a post-doctoral research fellow at Åbo Akademi University, Finland. He was also a lecturer and assistant professor at the department of computer science and engineering at the Khulna University of Engineering & Technology (KUET), Bangladesh during 2003-2012. His research areas include but not limited to human-computer interaction (HCI), humanitarian technology, health informatics, military information systems, information systems usability, and computer semiotics. He is the author of more than 80 peer-reviewed publications in International journals and conferences. Mr. Islam is a member of The Institution of Engineers, Bangladesh (IEB).

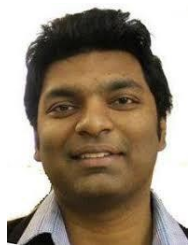

DR. A.K.M. NAJMUL ISLAM is an Adjunct Professor at Tampere University, Finland. He is a Scientist at LUT University, Finland. He also works as University Research Fellow at the Department of Future Technologies, University of Turku, Finland. Dr. Islam holds a PhD (Information Systems) from the University of Turku, Finland and an M.Sc. (Eng.) from Tampere University of Technology, Finland. He has 85+ publications. His research focuses on Human Centered Computing. His research has been published in top outlets such as IEEE Access, European Journal of Information Systems, Information Systems Journal, Journal of Strategic Information Systems, Technological Forecasting and Social Change, Computers in Human Behavior, Internet Research, Computers & Education, Journal of Medical Internet Research, Information Technology & People, Telematics & Informatics, Journal of Retailing and Consumer Research, Communications of the AIS, Journal of Information Systems Education, AIS Transaction on Human-Computer Interaction, and Behaviour & Information Technology, among others.

• • •